\documentstyle [aps,prc,preprint]{revtex}
\begin {document}
\draft

\title
{Role of correlated two--pion exchange in ${\bf K^+ N}$ scattering}

\author
{M. Hoffmann$^{1),2)}$, J.W. Durso$^{3)}$, K. Holinde$^{2),4)}$,
B.C. Pearce$^{5)}$, and J. Speth$^{2)}$}

\address
{1) Dept. of Physics, State University of New York, Stony Brook,
    N.Y. 11794, USA \\
 2) Institut f\"ur Kernphysik, Forschungszentrum J\"ulich,
    D-52425 J\"ulich, Germany \\
 3) Dept. of Physics, Mount Holyoke College, MA 01075, USA \\
 4) Departamento de Fisica Te\'orica and IFIC, Centro Mixto Universidad
    de Valencia--CSIC, E--46100 Burjassot (Valencia), Spain \\
 5) Dept. of Physics and Math. Physics, The University of Adelaide,
    Adelaide 5005, Australia}

\maketitle

\begin{abstract}
A dynamical model for S-- and P--wave correlated $2 \pi$ (and
$K \bar K$) exchange between a kaon and a nucleon is presented, starting
from corresponding $N \bar N \rightarrow K \bar K$ amplitudes in the
pseudophysical region, which have been constructed from nucleon,
$\Delta$--isobar and hyperon ($\Lambda$, $\Sigma$) exchange Born terms
and a realistic meson exchange model of the
$\pi \pi \rightarrow K \bar K$ and $K \bar K \rightarrow  K \bar K$
amplitude. The contribution in the s--channel is then obtained by
performing a dispersion relation over the unitarity cut. In the
$\rho$--channel, considerable ambiguities exist, depending on how the
dispersion integral is performed. Our model, supplemented by short range
interaction terms, is able to describe empirical $K^+ N$ data below pion
production threshold in a satisfactory way.
\end{abstract}

\pacs{ }

\section* {1. Introduction}

For quite some time, kaons have attracted the attention of nuclear
physicists. The reason is that strangeness, a quantum number conserved
in strong interactions, attributes a special role to the kaons among the
possible projectiles for investigating nuclear structure.
Kaons have two properties which make them unique tools. Firstly, they
can transfer a new degree of freedom to the nucleus, and secondly, in
contrast to pions they come in two forms, kaons ($K$) and antikaons
($\overline K$) which differ substantially in their interaction with the
nucleus. Kaons with their quark content $u \bar s$ ($d \bar s$) have
strangeness S=1; however, nuclear states involving strangeness can only
contain hyperons ($ \Lambda$, $ \Sigma$) which have the same strangeness
(S=--1) as antikaons (quark content $\bar u s$ or $\bar d s$).
Therefore, the interaction between kaons and nuclei is rather weak,
which is demonstrated, {\em e.g.}, by the large $K^+$ nuclear mean free
path of about 5--7 fm for $p_{lab} \le 0.8 {\rm GeV/c}$. Consequently,
the $K^+$ meson is a suitable probe for investigating the interior
region of nuclei. On the other hand, antikaons have S=--1 and their
absorption in nuclei can easily produce hypernuclei containing
$\Lambda$-- or $\Sigma$--hyperons. Since such processes occur with
sizeable reaction probability the region of investigation is probably
restricted to the nuclear surface.

The successful use of kaons in nuclear structure requires the precise
knowledge of the interaction mechanism of kaons with nuclei. Each
uncertainty in the theoretical description necessarily leads to
uncertainties in the interpretation of empirical results. Since each
theoretical model for the kaon--nucleus interaction starts from the
free kaon--nucleon interaction and then adds medium modifications in
one way or another, a precise knowledge of the free interaction is
absolutely essential.

Recently \cite{Bue} we have presented a meson exchange model for the
$K^+ N$ interaction, which provides a reasonable description of the
empirical $K^+ N$ scattering data for laboratory momenta smaller than
0.8 GeV/c. For model B of Ref. \cite{Bue} the diagrams included are
shown in Figs.~1(a) and 1(b); they have been evaluated in time--ordered
perturbation theory. By $\sigma_{\rm rep}$ we denote a very short
ranged, phenomenological, repulsive contribution, which has the
analytical form of scalar $\sigma$--exchange with opposite sign and with
an exchanged mass of 1.2 GeV. This additional repulsion is required if
the $\omega$--coupling  constants $g_{N N \omega}$, $g_{K K \omega}$ are
restricted to their SU(6) value.

An important ingredient is the contribution arising from $\sigma$--
and $\rho$--exchange: $\sigma$--exchange provides the dominant part of
the intermediate range attraction and $\rho$--exchange determines,
to a large part, the isospin dependence of the interaction. Despite
their importance, however, these pieces have been treated so far in a
very rough way. In both cases a sharp mass has been used which means
their appreciable decay widths have not been taken into account. The
$\sigma$ meson is a fictitious particle not observed in nature so both
its mass and coupling constant are therefore free parameters. In the
case of the $\rho$ meson the coupling strength is obtained as a product
of a coupling constant at the $NN \rho$ vertex, which is taken
from the Bonn potential \cite{MHoE} and the coupling constant at the
$KK \rho$ vertex, which is calculated via SU(3) relations from the
(empirically known) $\pi\pi\rho$ coupling constant. Whether this
procedure provides a reliable result is doubtful since i) the
$\rho$--meson coupling constants used in the Bonn potential are
questionable \cite{Kim} and ii) SU(3) relations are not necessarily
valid for unstable particles.

In this simple model, $\sigma$-- and $\rho$--exchange essentially
stand for the correlated $2\pi$--exchange contribution in the $J^P=0^+$
($\sigma$--) and $J^P=1^-$ ($\rho$--) channel, as illustrated in
Fig.~1(c). The purpose of the present paper is to derive this
contribution starting from a microscopic model for the t--channel
reaction $N \overline N \rightarrow K \overline K$ with $\pi\pi$ (and
$K \overline K$) intermediate states and using a dispersion
relation over the unitarity cut. This realistic model of (effective)
$\sigma$-- and $\rho$--exchange is then used to reconstruct an extended
meson exchange model for $KN$ scattering.

Such a microscopic description of correlated $2\pi$--exchange is
essential not only for an adequate judgment of the quantitative role of
meson exchange in free $K^+N$ scattering, but also for the calculation
of kaon--nucleus scattering processes. Effects of medium modifications
of meson masses \cite{NN} inevitably require an explicit, realistic
model for correlated $2\pi$--exchange.

In Sect.2 we outline the basic formalism. Sect.3 contains the essential
features of our model for the $N \overline N \rightarrow K \overline K$
transition. In Sect.4 we present the results for the correlated $2 \pi$
contribution in $KN$ scattering, in terms of suitably defined effective
coupling constants. Furthermore, we compare $KN$ phase shifts and
observables derived from the extended model with those obtained before.
Sect.5 contains some concluding remarks.

\section*{2. Formalism}

In this section we outline the formalism which we use in order to derive
the correlated $2 \pi$--exchange contribution for the $KN$ interaction.
The procedure is similar to that which was used in $\pi N$ scattering
\cite{SchuetzA}.

\subsection*{2.1 ${\bf KN \rightarrow KN}$ and
${\bf N \bar N \rightarrow K \bar K}$ amplitudes}
The scattering amplitude $T$ is related to the $S$ matrix by
\begin {equation}
S_{fi} = \delta_{fi} - i (2 \pi)^{-2} \delta^{(4)} (P_f - P_i)
  \left( \frac {m_N} {E_p} \frac {m_N} {E_{p'}} \right) ^{\frac {1}
  {2}}
  (2 \omega_q 2 \omega_{q'})^{- \frac {1} {2}} \ T_{fi} \, ,
\end {equation}
where $P_i$ ($P_f$) is the total four--momentum in the initial (final)
state, $E_p \equiv (\vec p \> ^2 + m_N^2)^{\frac{1}{2}}$, and
$\omega_q \equiv (\vec q \> ^2 + m_K^2)^{\frac{1}{2}}$ with $m_N$
($m_K$) the nucleon (kaon) mass.

In the s--channel ($KN \rightarrow KN$) $T$ can be written as
\begin {equation}
T_s(p',q',p,q) = \bar u(\vec p \>', \lambda') \xi^{\dagger} (\mu')
 \zeta^{\dagger} (\beta) \hat T(s,t) u(\vec p, \lambda) \xi (\mu)
 \zeta (\alpha) \; .
\label {s-chan}
\end {equation}
Here, the Dirac spinor $u(\vec p, \lambda)$ with the normalization
$\bar u u = 1$ describes a nucleon with helicity $\lambda$ and
three--momentum $\vec p$, while $\xi$ ($\zeta$) is the isospin wave
function of a nucleon (kaon). The operator $ \hat T$ acts in spin and
isospin space and depends on the two independent Mandelstam variables
$s \equiv (p+q)^2 = (p'+q')^2$ and $t \equiv (p'-p)^2 = (q-q')^2$. The
third variable $u$ is related to $s$ and $t$ by
$s+t+u = 2 m_N^2 + 2 m_K^2$. The scattering operator $\hat T$ has the
following isospin structure:
\begin {equation}
\hat T (s,t) = 3 \hat T^{(+)} {\bf 1} + 2 \hat T^{(-)} \vec \tau_N
                                                 \cdot \vec \tau_K
\, ,
\label {isospin}
\end {equation}
where $\vec \tau_N$ ($\vec \tau_K$) is the isospin operator for the
nucleon (kaon) and
\begin {equation}
\hat T^{(\pm)} (s,t) = - [A^{(\pm)}(s,t) I_4 +
Q \! \! \! \! \slash B^{(\pm)}(s,t)]
\, ,
\label {t-chan}
\end {equation}
with $Q \! \! \! \! \slash \equiv \gamma^{\mu} Q_{\mu}$,
$Q \equiv \frac {1} {2} (q' + q)$, and $I_4$ being the four--dimensional
unit matrix.

The corresponding t--channel ($N \bar N \rightarrow K \bar K$) amplitude
is given by
\begin {equation}
T_t(q',\bar q'; \bar p, p) = \bar v(\vec {\bar p}, \bar \lambda)
\bar \xi^{\dagger} (\bar \mu)
 \zeta^{\dagger} (\beta) \hat T(s,t) u(\vec p, \lambda) \xi (\mu)
 \zeta (\alpha) \, .
\end {equation}
with $\bar p \equiv -p'$, $\bar q' \equiv -q$ and
$\bar v$ ($ \bar \xi$)
the Dirac spinor (isospin state) of an antinucleon. The Mandelstam
hypothesis states now that $\hat T$ (and therefore $A^{(\pm)}$,
$B^{(\pm)}$ in Eq.(\ref{t-chan})) is the same function of s and t as in
Eq.(\ref{s-chan}), but in a different kinematical domain, {\em i.e.} for
$s=(p-\bar q')^2$, $t=(\bar p +p)^2$, and
$Q=\frac {1} {2} (q'- \bar q')$.

\subsection*{2.2 Spectral functions}
In order to isolate the $\sigma$ and $\rho$ contribution we have to
perform a partial wave decomposition of the amplitudes in the
t--channel:
\begin {equation}
A^{(\pm)} (s,t) \;  = \;
     \sum_J (J + {1 \over 2}) \, P_J(x) \, A_J^{(\pm)}(t)
\end{equation}
(the same for $B^{(\pm)}$). The $P_J(x)$ are the Legendre functions and
$x \equiv cos \theta_t$. The scattering angle in the t--channel,
$\theta_t$, can be expressed in terms of s and t:
\begin {equation}
x \; = \; \frac
    {s \; + \; \frac {1}{2} \, t \> -
    \> m_N^2 \> -\> m_K^2}
    {2 \sqrt {\frac{1}{4} t \> - \> m_N^2} \;
              \sqrt {\frac {1}{4} t \; - \; m_K^2}}   \; .
\end {equation}
Conservation of parity and G--parity demands that the sum of spin and
isospin must be even in case of a $2 \pi$ intermediate state.
Therefore $A_J^{(-)}$, $B_J^{(+)}$ ($A_J^{(+)}$, $B_J^{(-)}$) will
vanish for even (odd) $J$ if we only consider $2 \pi$ intermediate
states.
This argument does not hold for an intermediate $K \bar K$ state; there
$A_J^{(+)}$, $B_J^{(-)}$ ($A_J^{(-)}$, $B_J^{(+)}$) are also possible
for odd (even) $J$. However in our model these amplitudes turn out to
be negligibly small and can be safely neglected. Then the isospin index
$(\pm)$ can be suppressed since $J$ determines uniquely the isospin
state.

Since the $A$ and $B$ amplitudes contain kinematical singularities, one
has to define new amplitudes $f^J_{\pm}$, which are free of these
singularities \cite{FF}. Here the index  $\pm$ denotes the helicity of
the $N \bar N$ state:
$\lambda = \pm \frac {1}{2}$, $\bar \lambda = \frac {1}{2}$.
($f^{J=0}_-(t) = 0$.)
In terms of these amplitudes $A$ and $B$ can be written as
\begin {eqnarray}
A^{(\pm)} (s,t) \; & = & \; \frac {8 \pi}{p_t^2}
 \sum_J (J + \frac {1}{2}) \;
 (p_t q_t)^J
 \left \{ \frac {m_N}{\sqrt{J (J + 1)}} \, x \, P'_J(x) \, f^J_-(t)
                                \; -\; P_J(x) \, f^J_+(t) \right \}
         \nonumber \\
B^{(\pm)} (s,t) \; & = & \; 8 \pi
 \sum_J (J + \frac {1}{2}) \;
 \frac {(p_t q_t)^{J-1}}{\sqrt{J (J + 1)}} \, P'_J(x) \, f^J_-(t) \; ,
\label {AB-amp}
\end {eqnarray}
with $P'_J(x) = \frac {d}{dx} P_J(x)$ and
$p_t= \vert \vec p \> \! \vert$ ($q_t=\vert \vec q \> \! \vert$)
in the c.m. system of the t--channel process.
The $\sigma$-- ($\rho$--) exchange contribution, defined as the
correlated $2 \pi$--exchange in the scalar (vector) t--channel, is
identified as the $J=0$ ($J=1$) term of Eq.(\ref {AB-amp}).

One now can perform the analytic continuation of the $f$ amplitudes to
physical t--values in the s--channel ($t \le 0$), which requires
knowledge of the cut structure in the complex {\em t} plane.
The right hand (unitarity) cut runs from $4 m_{\pi}^2$ to $\infty$,
whereas the left hand cut, determined by $\Lambda$ exchange runs from
$-\infty$ to
$t_{\mbox{LH}} \equiv 4m_K^2 - \frac{(m_K^2+m_{\Lambda}^2-m_N^2)^2}
{m_{\Lambda}^2} > 4m_{\pi}^2$.
This means that the cuts overlap. In this overlap region the baryon
exchange Born term of the reaction $N \bar N \rightarrow K \bar K$
has an imaginary part. As we are only interested in the correlated
$2 \pi$ contribution (which we call $\tilde f^J_{\pm}$ in the following)
we perform the dispersion relation for $f^J_{\pm}$ only over the right
hand cut, leaving out the Born term. Then we get for $\sigma$--exchange
the following contributions to the invariant amplitudes in the
s--channel:
\begin {eqnarray}
A_{\sigma}^{(+)}(s,t) & = &
 - \frac {4 \pi} {p_t^2} \tilde f^0_+ (t) \; = \;
-16 \int_{4 m_\pi^2}^{\infty} \frac {{\rm I\;\!\!m} \,
    \tilde f^0_+ (t')dt'} {(t' - t) \, (t' - 4 m_N^2)}
\nonumber \\
B_{\sigma}^{(+)}(s,t) & = &  0 \; .
\label {AB-sigma}
\end {eqnarray}
(For convergence reasons, see Ref. \cite {FF}, the dispersion relation
has to be performed for $\frac {\tilde f^0_+(t)}{p_t^2}$,
$p_t^2 = \frac {t}{4} - m_N^2$.)
Similarly we obtain for $\rho$--exchange
\begin {eqnarray}
A^{(-)}_\rho(s,t) \; & = & \;
12 \pi \, \frac {q_t}{p_t} \, x \;
\left (\frac {m_N}{\sqrt 2} \, \tilde f^1_-(t) \; - \; \tilde f^1_+(t)
                                                        \right )
 \nonumber \\
 & = & \; 12
 \frac {s + \frac {1}{2} t - m_N^2 -  m_K^2}
   {t \; - \; 4 m_N^2}
   \Bigg( \sqrt 2 m_N
    \int_{4 m_\pi^2}^{\infty} \frac {{\rm I\;\!\!m} \,
     \tilde f^1_- (t')}  {t' - t} \, dt' \; - \;  2
    \int_{4 m_\pi^2}^{\infty} \frac {{\rm I\;\!\!m} \,
    \tilde f^1_+ (t')}   {t' - t} \, dt' \Bigg)
  \nonumber \\
B^{(-)}_\rho(s,t) \; & = & \;
6 \sqrt 2 \, \pi \, \tilde f^1_-(t)
\; = \;  6 \sqrt 2
    \int_{4 m_\pi^2}^{\infty} \frac {{\rm I\!m} \, \tilde f^1_- (t')}
      {t' - t} \, dt'  \; .
\label {AB-rhof}
\end {eqnarray}
Here s and t have to assume physical values of the s--channel,
{\em i.e.}, $s \ge (m_N + m_K)^2$ and $t \le 0$.
These amplitudes can be interpreted as meson exchange potentials
for which the meson mass $\sqrt {t'}$ is distributed over the range
from $2 m_{\pi}$ to $\infty$. The corresponding coupling constants
depend on the mass; they are proportional to the spectral functions
${\rm I\;\!\!m} \, \tilde f^0_+ (t')$,
${\rm I\;\!\!m} \, \tilde f^1_{\pm} (t')$. Therefore, these spectral
functions determine the dynamical behaviour of the exchanges,
{\em i.e.}, they characterize the strength as well as the range of
these potentials.

Concerning $\rho$--exchange, it was pointed out in Refs.
\cite{SchuetzA,SchuetzB} that there is a considerable uncertainty in the
results. The reason is that Eq.(\ref {AB-rhof}) disperses the helicity
amplitudes directly; alternatively, one (\cite{HoehPie}) can first
construct combinations $\tilde \Gamma_{1,2}(t)$ corresponding to vector
($\tilde \Gamma_1$) and tensor ($\tilde \Gamma_2$) coupling amplitudes,
where
\begin {eqnarray}
\tilde \Gamma_1(t) & =
 \displaystyle{- \frac {m_N}{\frac {t}{4} - m^2_N}} & \left \{
\tilde f^1_+(t) - \frac {t} {4 \sqrt 2 m_N} \tilde f^1_-(t)
\right \}
      \nonumber \\
\tilde \Gamma_2(t) & =
 \displaystyle{+ \frac {m_N}{\frac {t}{4} - m^2_N}} &  \left \{
 \tilde  f^1_+(t) - \frac {m_N} {\sqrt 2} \tilde f^1_-(t)
  \right \} \; ,
\label {Gamma}
\end {eqnarray}
and then perform the dispersion integral, which yields
\begin {eqnarray}
A^{(-)}_{\rho}(s,t) \; & = & \;
         - 12 \pi \frac {q_t p_t x}{m_N} \tilde \Gamma_2(t)
  \nonumber \\
 & = & \; - 6
 \frac {s + \frac {1}{2} t - m_N^2 -  m_K^2} {m_N}
    \int_{4 m_\pi^2}^{\infty} \frac {{\rm I\;\!\!m} \,
    \tilde \Gamma_2(t')} {t' - t} \, dt'
      \nonumber \\
B^{(-)}_{\rho}(s,t) \; & = & \;
         - 12 \pi (\tilde \Gamma_1(t) + \tilde \Gamma_2(t))
      \nonumber \\
 & = & \; - 12
   \Bigg(
    \int_{4 m_\pi^2}^{\infty} \frac {{\rm I\;\!\!m} \,
    \tilde \Gamma_1(t')} {t' - t} \, dt' \; + \;
    \int_{4 m_\pi^2}^{\infty} \frac {{\rm I\;\!\!m} \,
    \tilde \Gamma_2(t')} {t' - t} \, dt' \Bigg)  \; .
\label {AB-rhoG}
\end {eqnarray}
Both methods would be equivalent if the dispersion integrals could be
performed over both the complete left hand and unitarity cut. However,
$\rho$--exchange is customarily defined via a dispersion integral over
the unitarity cut only. Indeed, the additional t--dependence in
$\tilde \Gamma_i$ apart from the t--dependence provided by the helicity
amplitudes $\tilde f^J_{\pm}$ causes the results to be quite
different, as will be discussed later.

\subsection*{2.3 Correlated $2 \pi$--exchange potentials}
Based on Eqs. (\ref {s-chan}) -- (\ref {t-chan}) the correlated
$2 \pi$--exchange contributions to the $KN$ scattering amplitude can
then be written as
\begin {eqnarray}
T_s^{(\sigma)} & = &
  - 3 \, \bar u(\vec p \> ', \lambda') A_{\sigma}^{(+)}(t)
  u(\vec p, \lambda) {\bf 1} \; ,
     \nonumber \\
T_s^{(\rho)} & = &
  - 2 \, \{\bar u(\vec p \> ', \lambda')
  [A^{(-)}_{\rho}(s,t) + Q \! \! \! \! \slash B^{(-)}_{\rho}(t)]
  u(\vec p, \lambda) \} \vec \tau_N \cdot \vec \tau_K \; ,
\label {T-sigrho}
\end {eqnarray}
omitting isospin states for convenience.
In the c.m. system of the $KN \rightarrow KN$ reaction
$s = (E_p + \omega_p)^2$, $t = -(\vec p \>' - \vec p)^2$, and
$Q = (\omega_p, - \frac {1}{2} (\vec p \>' + \vec p))$. Since we will
treat $T^{(\sigma, \rho)}_s$ later as potentials to be iterated in the
scattering equation belonging to time--ordered perturbation theory, we
apply an off-shell extrapolation of the dispersion relations in the
following way: We first replace the denominator $t'-t$ in analogy to the
time--ordered propagator
\begin {eqnarray}
\frac {1}{t - t'} \rightarrow  \frac {1}{2 \omega_r} & \Bigg ( &
    \frac {1}{Z - \omega_r - E_p -\omega_{p'}}
     \nonumber \\
   & + &  \frac {1}{Z - \omega_r - E_{p'} -\omega_p} \Bigg )
\label{prop}
\end {eqnarray}
($\omega_r \equiv [t' + (\vec p \> ' -\vec p)^2]^{\frac {1}{2}}$, $Z =
E_{p_{on}} + \omega_{p_{on}}$ is the total c.m.\ energy), which is
motivated by the fact that the dispersion integral sums over exchanges
of particles with mass $\sqrt{t'}$. For the additional t dependence in
Eqs.(\ref {AB-rhof},\ref {AB-rhoG}) we keep
$t =- (\vec p \> ' - \vec p)^2$ and
$s \equiv (E_p + \omega_p)(E_{p'} + \omega_{p'})$. Note that this does
not change the on--shell result.

In addition we have to add phenomenological cutoffs in order to generate
sufficient convergence in the scattering equation. Again we interpret
the correlated $2 \pi$ potentials as generated by exchange of particles
with mass $\sqrt {t'}$ and define a formfactor
\begin {equation}
F(t) = \frac {\Lambda^2_{\sigma,\rho} - t'}
       {\Lambda^2_{\sigma,\rho} - t}   \; ,
\label{formfac}
\end {equation}
which is squared under the dispersion integral. This is analogous to the
monopole type form factor we use at each vertex of the ordinary meson
and baryon exchange potentials.  We use
$\Lambda_{\sigma,\rho} = 1850 (2400)$ MeV for the model where we perform
the dispersion integral for $\tilde f^1_{\pm}(t)$
($\tilde \Gamma_i(t)$). One should realize that this procedure modifies
the original on--shell result somewhat. However, it lies in the range of
uncertainties which are inherent in the whole procedure, which are
discussed further in Sect.4.

With the above extensions the amplitudes in Eq.(\ref{T-sigrho}) have a
well defined off--shell behaviour with a sufficient fall--off for
high momenta. Corresponding potential matrix elements,
$<\!\vec p \>' \lambda'|V(Z)|\vec p \lambda\!>$,
acquire an additional factor
$\kappa= \frac {1}{(2 \pi)^3} \sqrt {\frac {m_N}{E_p} \frac {m_N}{E_p'}}
\sqrt {\frac {1} {2 \omega_p} \frac {1} {2 \omega_{p'}}}$, {\em i.e.}
$<\!\vec p \>' \lambda'|V_{\sigma, \rho}(Z)|\vec p \lambda\!> =
\kappa
<\!\vec p \>' \lambda'|T^{(\sigma, \rho)}_s(Z)|\vec p \lambda\!>$.
Unitarization then leads to the scattering amplitude, {\em i.e.}
\begin {eqnarray}
<\vec p \>' \lambda'|T(Z)|\vec p \lambda> = & &
<\vec p \>' \lambda'|V(Z)|\vec p \lambda> +
 \\ \nonumber
& & \sum_{\lambda''} \! \int d^3 p''\!
<\vec p \>' \lambda'|V(Z)|\vec p \>'' \lambda''>
\frac {1} {Z-E_{p''}-\omega_{p''}+i \epsilon} \!
<\vec p \>'' \lambda''|T(Z)|\vec p \lambda> \; ,
\end {eqnarray}
where $V$ contains contributions from the diagrams shown in Fig.~1,
except that now the $\sigma$ and $\rho$ exchange potentials are replaced
by the correlated $2\pi$ exchange potentials discussed here.

\section*{3. Microscopic model for the
${\bf N \bar N \rightarrow K \bar K}$ process}

In the last section we have outlined a method of obtaining the
correlated $2 \pi$--exchange contribution to the $KN \rightarrow KN$
scattering amplitude from the $N \bar N \rightarrow K \bar K$ partial
wave helicity amplitudes $f^J_{\pm}$. Since, unlike for the
$N \bar N \rightarrow 2 \pi$ case (Ref. \cite{SchuetzA}), we cannot rely
on quasiempirical information, we have to provide a field--theoretic
model for the $N \bar N \rightarrow K \bar K$ amplitudes. Anyhow, such a
dynamical model has definite advantages when medium modifications of the
$KN$ interaction are considered since it facilitates future
investigation of not only possible medium effects due to changes in the
kaon and nucleon propagators, but also in the
$N \bar N \rightarrow K \bar K$ interaction itself.

We will generate the amplitude for the process of Fig.~1(c) in the
t--channel by solving the scattering equation in the
Blankenbecler--Sugar (BbS) \cite{BbS} reduction scheme:
\begin{equation}
T_{N \bar N \rightarrow K \bar K} \, = \,
V_{N \bar N \rightarrow K \bar K} \, + \,
\sum_{aa = \pi \pi,K \bar K}
T_{aa \rightarrow K \bar K} \ g_{aa} \
V_{N \bar N \rightarrow aa}  \; ,
\label{tchanT}
\end{equation}
where
\begin{equation}
T_{aa \rightarrow K \bar K} = V_{aa \rightarrow K \bar K}
  + \sum_{bb = \pi \pi,K \bar K}
T_{bb \rightarrow K \bar K} \ g_{bb} \
V_{aa \rightarrow bb} \; .
\end{equation}
Here $V_{N \bar N, aa}$ is the transition interaction from
$N \bar N$ to $aa= \pi \pi, {K \bar K}$,
$T_{aa, K \bar K}$ are transition amplitudes from $\pi \pi$ and
$K \bar K$ to $K \bar K$, and $g_{aa}$ is the free two--particle
Green's function for the {\em aa} intermediate state. The ingredients of
the dynamical model for the transition interactions
$V_{N \bar N, \pi \pi}$ and $V_{N \bar N, K \bar K}$ are shown in
Fig.~2. The potential $V_{N \bar N, \pi\pi}$ ($V_{N \bar N, K \bar K}$)
consists of $N$ and $\Delta$ ($\Lambda$ and $\Sigma$) exchange terms
plus $\rho$ meson pole diagrams. $T_{\pi \pi, K \bar K}$ and
$T_{K \bar K, K \bar K}$ are obtained from the driving terms shown in
Fig.~3. Such a model involving the coupled channels $\pi \pi$ and
$K \bar K$ was constructed by our group \cite{Loh} based on
time--ordered perturbation theory. Here, as in a recent study of
$\pi N$ scattering \cite{SchuetzB}, we use a model with essentially the
same physical input, which alternatively uses the BbS scheme. The
description of the data turns out to be as successful as in
Ref. \cite{Loh}. For more details, the reader is referred to
\cite{SchuetzB}.

We stress that all parameters are predetermined: $T_{\pi \pi, K \bar K}$
and $T_{K \bar K, K \bar K}$, through the coupled channel calculation,
are fixed by $\pi \pi$ while both transition potentials are determined
by the quasiempirical $N \bar N \rightarrow 2 \pi$ information, cf.
Ref. \cite{SchuetzB}.

In the c.m. system and in helicity representation Eq.(\ref{tchanT})
after a partial wave expansion becomes
\begin {eqnarray}
& &<00|T^J_{N \bar N \rightarrow K \bar K}(q,p;t)|
                  \lambda_N \lambda_{\bar N}> =
   <00|VJ_{N \bar N \rightarrow K \bar K}(q,p;t)|
                  \lambda_N \lambda_{\bar N}>
 \nonumber   \\
& & + \sum_{aa} \int_0^\infty dk k^2 \frac
{<00|T^J_{aa \rightarrow K \bar K}(q,k;t)|00>
 <00|V^J_{N \bar N \rightarrow aa}(q,k;t)|\lambda_N \lambda_{\bar N}>}
{(2 \pi)^3 \ \omega_a(k) \ (t-\omega_a^2(k))}
\end {eqnarray}
with
\begin {equation}
\omega_a(k) = \sqrt{k^2 +m_a^2}  \; .
\end {equation}
The $N \bar N \rightarrow K \bar K$ on--shell amplitudes are related
to the helicity amplitudes $f^J_{\pm}$ via
\begin {eqnarray}
f^J_+(t) &=& \frac {p_{on} m_N} {4 (2 \pi)^2 (p_{on}q_{on})^J}
 <00|T^J_{N \bar N \rightarrow K \bar K}(q_{on},p_{on};t)|
                                          \frac {1}{2} \frac {1}{2}>
 \nonumber \\
f^J_-(t) &=& - \frac {p_{on} m_N} {2 (2 \pi)^2 \sqrt{t}(p_{on}q_{on})^J}
 <00|T^J_{N \bar N \rightarrow K \bar K}(q_{on},p_{on};t)|
                                          \frac {1}{2} (-\frac {1}{2})>
\end {eqnarray}
with
\begin {eqnarray}
q_{on} &=& \sqrt{\frac {t}{4} - m_K^2}
 \nonumber \\
p_{on} &=& \sqrt{\frac {t}{4} - m_N^2}   \; .
\end {eqnarray}

Fig.~4 shows the results for ${\rm Im} \tilde f^0_+$ and
${\rm Im} \tilde f^1_{\pm}$ in the pseudophysical region
($t > 4 m_{\pi}^2$), needed as input for Eqs.(\ref{AB-sigma},
\ref{AB-rhof}, \ref{AB-rhoG}). As expected, the spectral function in
the $\rho$--channel shows a resonant structure with a maximum at about
the $\rho$--mass. In the $\sigma$--channel the spectral function is much
broader than for the $\rho$; compared to the
$N \bar N \rightarrow2 \pi$ case (Ref. \cite{SchuetzA}) it is weaker and
the peak is shifted somewhat to a higher mass (see also the discussion
later). Furthermore, the inclusion of intermediate $K \bar K$ states
leads to a sizable enhancement of ${\rm Im} \tilde f^0_+$ whereas its
effect is negligible in the $\rho$--channel. This is due to the fact
that the $K \bar K$ interaction is weak in the vector but strong in the
scalar channel. In fact, as discussed in Refs.\cite{Loh} and
\cite{Janssen}, the $K\bar K$ interaction generates a $K\bar K$ bound
state, the $f_{0}(975)$ meson. This state clearly has a strong effect on
the shape of ${\rm Im} \tilde f^0_+$.

\section*{4. Results}

\subsection*{4.1 Effective coupling constants}
Based on the spectral functions of the last section we can now evaluate,
in a first step, the (on--shell) invariant amplitudes with the help of
Eqs.(\ref{AB-sigma}, \ref{AB-rhof}, \ref{AB-rhoG}). In practice, the
integrals have been evaluated up to $t_c = 120 m_{\pi}^2$,
{\em i.e.}, in a region in which the dynamical model can be trusted.

It is instructive to parametrize the result by sharp mass $\sigma$--
and $\rho$--exchange with appropriate t--dependent coupling constants
which can be compared with those used in our former model. Let us start
with the $\sigma$--channel. From the Lagrangians used in Ref. \cite{Bue}
\begin {eqnarray}
{\cal L}_{NN \sigma} & = &g_{NN \sigma} \, \overline \psi_N \, \psi_N \,
                                          \phi_{\sigma}
           \nonumber \\
{\cal L}_{KK \sigma} & = &g_{KK \sigma} \, m_K \, \phi_K \, \phi_K \,
                                           \phi_{\sigma}   \; ,
\end {eqnarray}
we get for the invariant amplitude arising from $\sigma$--exchange:
\begin {equation}
A^{(+)\>'}_{\sigma}(t)=
       -\frac{2 g_{\sigma} \, m_K \,}{m^2_{\sigma} - t} \; ,
\end {equation}
with $g_{\sigma} \equiv g_{NN \sigma}g_{KK \sigma}$ .
If we now parametrize the result of our correlated $2 \pi$--exchange
potential (Eq.(\ref {AB-sigma})) in this form (allowing $g_{\sigma}$
to be t--dependent) we will get for the effective coupling constant
\begin {equation}
\frac {g_{\sigma}(t)}{4 \pi} = - \frac {1}{8 \pi m_K} \,
                A^{(+)}_{\sigma} (t) \, (m^2_{\sigma} - t)  \; ,
\label {coup-sig}
\end {equation}
The result (with $m_{\sigma} = 0.6$ GeV) is shown in Fig.~5, together
with the result of an updated model of Ref. \cite{Bue}, hereafter
referred to as model I, based on the diagrams in Figs.~1(a) and 1(b),
with the same value of $m_{\sigma}$. (Details of this model will be
given below.) Obviously our model predictions for the correlated
$2 \pi$--exchange in the scalar channel is in rough agreement with the
$\sigma$--strength used before, which was phenomenologically adjusted to
empirical $K^+N$ data. However, our new result has a sizable
t--dependence, which demonstrates clearly that it cannot be well
approximated by sharp mass $\sigma$--exchange with
$m_{\sigma} = 0.6$GeV. Since it grows with $-t$, the contribution is
shorter ranged, in complete consistency with the behaviour of the
spectral function in Fig.~4(a). Therefore, the effective mass exchanged
should be higher. Indeed, if we use $m_{\sigma} = 0.75$GeV in
Eq.(\ref{coup-sig}) the result for $g_{\sigma}(t)$, shown in the dashed
curve of Fig.~5, has almost no t--dependence. Note that the analogous
effective $\sigma$--mass for the $\pi N$ system \cite{SchuetzA} is
$m_{\sigma} = 0.55$GeV.

Let us now go to the $\rho$--channel. Starting from the Lagrangians
for sharp mass $\rho$--exchange
\begin {eqnarray}
{\cal L}_{NN \rho} & = & \overline \psi_N \, \{ g^V_{NN \rho} \,
                \gamma_{\mu} \vec \phi_{\rho}^{\mu} \, + \,
 \frac {1}{4 m_N} g^T_{NN \rho} \sigma_{\mu \nu} \times
    (\partial^{\mu} \vec \phi_{\rho}^{\nu} \, - \,
     \partial^{\nu} \vec \phi_{\rho}^{\mu}) \}
 \vec \tau \psi_N \,
           \nonumber \\
{\cal L}_{KK \rho} & = & g_{KK \rho} \,
    (\phi_K \vec \tau \partial^{\mu} \phi_K) \, (\vec \phi_{\rho})_{\mu}
\label{L-rho}
\end {eqnarray}
we get for the invariant amplitudes:
\begin {eqnarray}
A^{(-)\>'}_{\rho}(s,t) & = &
      - \frac {q_t p_t x}{m_N} \frac {2 g_T}{m_{\rho}^2 -t }
           \nonumber \\
B^{(-)\>'}_{\rho}(t) & = &
        \frac {2 (g_V + g_T)}{m_{\rho}^2 -t }  \; ,
\end {eqnarray}
with $g_V \equiv g^V_{NN \rho}g_{KK \rho}$ ,
$g_T \equiv g^T_{NN \rho}g_{KK \rho}$. Parametrizing again our
correlated $2 \pi$--exchange result (Eq.(\ref {AB-rhof}) resp.
Eq.(\ref {AB-rhoG})) in this form we obtain for the effective $\rho$
coupling constants
\begin {eqnarray}
\frac {g_V(t)}{4 \pi} & = & + \frac {1}{8 \pi} \,
    \left( \frac {m_N} {q_t p_t x} A^{(-)}_{\rho}(s,t) \, + \,
           B^{(-)}_{\rho}(t) \right) \, (m^2_{\rho} - t)
  \nonumber \\
\frac {g_T(t)}{4 \pi} & = & - \frac {1}{8 \pi} \,
       \frac {m_N} {q_t p_t x} \, A^{(-)}_{\rho}(s,t) \,
       (m^2_{\rho} - t)  \; .
\end {eqnarray}
(Note that according to Eqs.(\ref {AB-rhof}, \ref {AB-rhoG}) the
s-dependence drops out.)

The results for $g_V$ and $g_T$ based on $m_{\rho}=769$ MeV are shown in
Fig.~6, again together with values used in model I. First we observe
that there is a remarkable difference between the two alternatives for
doing the dispersion relation in the $\rho$--channel, {\em i.e.},
Eq.(\ref{AB-rhof}) on one hand and Eq.(\ref{AB-rhoG}) on the other: The
second method provides larger tensor ($g_T$) but smaller vector coupling
($g_V$). Consequently the ratio $g_T/g_V$ characterizing the tensor to
vector coupling ratio of the (effective) $\rho$ to the nucleon,
$g^T_{NN \rho}/g^V_{NN \rho}$ is much larger ($\sim 5$) for the second
than for the first choice ($\sim 2.5$). Note also that the results based
on the second choice are almost t--independent; therefore, the result
can be well identified with an exchange of an (effective) $\rho$ meson
with the empirical mass. On the other hand, the first method yields a
result with a non--negligible t--dependence, with opposite behaviour for
$g_V$ and $g_T$. Therefore, given the Lagrangians in Eq.(\ref{L-rho}),
a common mass cannot be assigned to the $\rho$--channel result derived
from the first choice.

The situation is similar to the $\pi N$ case (cf. Fig.~9 of
Ref. \cite{SchuetzB}). The only difference is that there all results
scale by a factor of (roughly) 3, in distinct disagreement with the
SU(3) value of 2 for the $g_{\pi \pi \rho}/g_{K K \rho}$ ratio. (Note
that there is an additional factor 2 due to the factor $\frac {1}{2}$ in
the Lagrangian used in \cite{SchuetzB}.) This discrepancy should be of
no surprise since there is no reason to expect that such effective
exchanges generated by correlated $2 \pi$ exchange should fulfill the
symmetry relations for exchanges of genuine particles.

In the sharp mass $\rho$--model I the $g_T/g_V$ ratio has been taken
from the Bonn $NN$ model potential \cite{MHoE} to be 6.1, only slightly
larger than the correlated result based on Eq.(\ref{AB-rhoG}). On the
other hand, the absolute values are much larger, by about a factor of 2.
Note however, that in the following calculation of $KN$ phase shifts and
observables this discrepancy in the physical t region is much reduced
since in this model a $\rho N N$ (and $\rho K K$) formfactor of
monopole type with a cutoff mass of only about twice the $\rho$--mass
are introduced, which suppress the $\rho$--potentials by about a factor
of 2 at t=0.

\subsection*{4.2 Model for KN scattering}
In this section we confront our correlated $2 \pi$--exchange model with
the experimental $K^+ N$ data. Our starting point is model B of
Ref. \cite{Bue} consisting of the diagrams shown in Figs.~1(a) and 1(b).
The baryon exchange diagrams in Fig.~1(b) are now treated in an improved
way. First, both time orderings are included instead of only one
(cf. Fig.~1(b) of Ref. \cite{Bue}). Second, pseudovector coupling is
used at the NYK vertex, {\em i.e.}
\begin {eqnarray}
{\cal L}_{N \Lambda K} & = &
  \frac {f_{N \Lambda K}}{m_K} \;
 (\overline{\psi}_{\Lambda}\,(x) \;
     \gamma^5 \gamma^\mu \; \psi_N\,(x) \; + \;
  \overline{\psi}_N\,(x) \;
     \gamma^5 \gamma^\mu \; \psi_{\Lambda}\,(x)) \; \partial ^\mu \,
     \phi_K \,(x) \; .
 \nonumber \\
{\cal L}_{N \Sigma K} & = &
  \frac {f_{N \Sigma K}}{m_K} \;
 (\vec {\overline{\psi}}_{\Sigma}\,(x) \;
     \gamma^5 \gamma^\mu \; \psi_N\,(x) \; + \;
  \overline{\psi}_N\,(x) \;
     \gamma^5 \gamma^\mu \; \vec \psi_{\Sigma}\,(x)) \;\;
     \vec \tau \, \partial ^\mu \phi_K \,(x) \; .
\end {eqnarray}
Third, in case of $Y^*$--exchange, an extended spin--3/2 propagator
is taken. Finally monopole formfactors (Eq.(2.17) of Ref. \cite{Bue})
are used throughout, except for $NY^*K$ and $N\Delta\rho$ vertices,
where a dipole formfactor is used. The new expressions for the potential
matrix elements are given in the appendix. In addition, we change
$g^2_{N \Delta \pi}$ to its experimental value
$\frac {g^2_{N \Delta \pi}} {4 \pi} = 0.36$, instead of its quark
model value used before. Correspondingly, via SU(6) relations,
$\frac {g^2_{N \Delta \rho}} {4 \pi}$ now becomes 32.95.

As in Ref. \cite{Bue}, most parameters (coupling constants, cutoff
masses) are predetermined: Coupling constants (with the exception of
$g^2_{N \Delta \pi}$ and $g^2_{N \Delta \rho}$) and cutoff masses
belonging to $NN$ and $N \Delta$ vertices are taken to be precisely the
same as those of the (full) Bonn $NN$ potential \cite{MHoE}. Coupling
constants at the vertices involving strange baryons ($g_{N \Lambda K}$,
$g_{N \Sigma K}$, $g_{N Y^* K}$) have been related by the assumption of
SU(6) symmetry to the empirical $NN \pi$ coupling, as in our
hyperon--nucleon model \cite{Holz}. The three--meson coupling constants
have been determined from the empirical $\pi \pi \rho$ coupling,
assuming the same symmetry scheme and ideal mixing. The value of
$g_{KK \sigma}$ and some cutoff masses have been slightly readjusted to
the empirical $K^+ N$ phase shifts below pion production threshold.
This defines our model I. The values of parameters used in this model
are given in Table 1; the phase shifts which it yields are shown in
the dotted curves of Fig.~7. Obviously, the model based on
phenomenological sharp mass $\sigma$-- and $\rho$--exchange provides a
fair description of the empirical situation, with slight improvements
in the $P_{03}$ and $P_{13}$ phase shifts compared to Ref. \cite{Bue}.

We now replace the sharp mass $\sigma$ and $\rho$ contributions by the
correlated $2 \pi$--exchange potentials based on Fig.~1(c) (model
I\negthinspace I, A=Eq.(\ref {AB-rhof}), B=Eq.(\ref {AB-rhoG}))
evaluated off--shell using Eq.(\ref{prop}) and including formfactors
(Eq.(\ref{formfac})). In order to avoid double counting we have then to
omit the box diagrams in Fig.~1(b) involving two pions since they are
already included in Fig.~1(c).

After a slight readjustment of some parameters, cf. Table 2, we obtain
phase shift results shown likewise in Fig.~7. As expected from
Figs.~5 and 6, some discrepancies occur between the various models. The
main point however is that $K^+ N$ interactions based on a microscopic
evaluation of correlated $2 \pi$ exchange are able to provide a
reasonable description of empirical phase shifts.

Since the existing phase shift analyses have large error bars and are,
in some cases, even contradictory it is instructive to examine the
experimental observables directly. Fig.~8 shows our model predictions
for the elastic cross sections in the relevant momentum range, while
Figs.~9--11 show the differential cross sections and polarizations at
some selected momenta. All models are in good agreement with
experimental data. Again, there are slight differences between the
various model results. The differential cross sections for $K^+p$
suggest an almost complete absence of P--waves, which is best realized
in the model involving correlated $2 \pi$--exchange evaluated according
to Eq.(\ref{AB-rhof}).

Finally, Table 3 presents scattering lengths $a^I_{\frac{1}{2} S}$
(I=0,1) and effective range $r^1_{\frac{1}{2} S}$ of our models, which
are in overall agreement with the empirical values.

\section*{5. Summary}

In this paper we have presented a microscopic model for correlated
$2 \pi$ (and $K \bar K$)--exchange between kaon and nucleon,
in the scalar--isoscalar ($\sigma$) and vector--isovector ($\rho$)
channels. We first constructed a model for the reaction
$N \bar N \rightarrow K \bar K$ with intermediate $2 \pi$ and
$K \bar K$ states, based on a transition in terms of baryon
($N, \Delta, \Lambda, \Sigma$) exchange and a realistic coupled channel
$\pi\pi\rightarrow\pi\pi$, $\pi\pi\rightarrow K\bar K$ and
$K\bar K\rightarrow K\bar K$ amplitude. The contribution in the
s--channel is then obtained by performing a dispersion relation over
the unitarity cut.

In the $\sigma$--channel, the result can be suitably represented by
an exchange of a scalar particle with a mass of 0.75 GeV. The strength
turns out to be in rough agreement with the strength of phenomenological
$\sigma$--exchange used before \cite{Bue}, which has been adjusted,
together with other diagrams, to empirical $K^+ N$ data.

In the $\rho$--channel, considerable ambiguities exist, of the same
structure as in the $\pi N$ case, depending on how the dispersion
integral is performed. In terms of effective coupling constants, the
results differ strongly from the values used before in our
phenomenological $\rho$--exchange amplitude, which were determined from
the Bonn $NN$ potential and SU(3) relations. However, the
$\rho$--amplitudes actually used are quite similar since the formfactor
applied in the phenomenological $\rho$--exchange of model I brings the
contribution close to the results of the dispersion--theoretic results.

This model for correlated $2 \pi$--exchange has been suitably
extrapolated off--shell, and supplemented by short range terms
(generated partly by conventional $\omega$--exchange) and box diagrams
involving $\pi \rho$-- and $\rho \rho$--exchange developed before. A
satisfactory description of the empirical situation is achieved, of the
same overall quality as obtained before using phenomenological sharp
mass $\sigma$-- and $\rho$--exchange.

Such an explicit model for correlated $2 \pi$--exchange has not only
conceptual advantages compared to a phenomenological treatment in terms
of $\sigma$--, $\rho$--exchange, but also offers the possibility to
study medium modifications of the $KN$ amplitude in a well--defined
way---a topic of high current interest.

\vskip1cm
\centerline {\bf ACKNOWLEDGMENTS}
One of the authors (K. H.) wishes to acknowledge the Ministerio da
Educacion y Ciencia for financial support (SAB94--0218) making his stay
in Valencia possible. BCP is grateful for the support of the
Australian Research Council. Part of the work has been done at the
Institut de Physique Nucleaire d'Universit\'e Claude Bernard Lyon I,
supported by the DAAD program PROCOPE. M. H. wants to thank Guy Chanfray
for his hospitality. This work was also partially supported by the DAAD
program HSPII/AUFE.

\appendix
\section{Potential matrix elements for baryon exchange}
For the baryon exchange diagrams we get the following potential
matrix elements:
\begin{eqnarray}
<\vec p \>' \lambda'|V_Y|\vec p \lambda> = & - & \kappa
   \frac {f^2_{NYK}} {m_K^2} q_{\mu} q'_{\nu}
        \bar u(\vec p \>',\lambda') \gamma^5 \, \gamma^{\mu}
               (p_r \! \! \llap {/} + m_r)
        \gamma^5 \, \gamma^{\nu}  u(\vec p,\lambda)
 \nonumber \\
& & \cdot \frac {1}{2 E_r}
\left ( {\frac {1} {Z - E_r - E_p- E_{p'}} \ + \
         \frac {1} {Z - E_r - \omega_q - \omega_{q'}} } \right )
\ F_Y(I) \; ,
\end{eqnarray}
where the isospin factors are
$F_{\Lambda} = \frac {1} {2} (1 + \vec \tau_1 \cdot \vec \tau_2)$
and
$F_{\Sigma} = \frac {1} {2} (3 - \vec \tau_1 \cdot \vec \tau_2)$.

In case of $Y^*$--exchange we get
\begin{eqnarray}
<\vec p \>' \lambda'|V_Y|\vec p \lambda> = & - & \kappa
\frac {f^2_{NY^*K}} {m_K^2}
    \bar u(\vec p \>',\lambda') q'_{\mu} P^{\mu \nu}(p_r) q_{\nu}
                           u(\vec p,\lambda) \ F_{Y^*}(I)
\nonumber \\
= & - & \kappa  \frac {f^2_{NY*K}} {m_K^2} q_{\mu} q'_{\nu}
    \bar u(\vec p \>',\lambda') (p_r \! \! \llap {/} + m_r)
\nonumber \\
& & \left \{
- g^{\mu \nu} \ + \ \frac {1} {3} \gamma^{\mu} \gamma^{\nu} \ + \
 \frac {2} {3 m_r^2} p_r^{\mu} p_r^{\nu} \ - \
 \frac {1} {3 m_r} (p_r^{\mu} \gamma^{\nu} - p_r^{\nu} \gamma^{\mu})
\right \}
  u(\vec p,\lambda)
\nonumber \\
& & \cdot \frac {1}{2 E_r}
\left ( {\frac {1} {Z - E_r - E_p- E_{p'}} \ + \
         \frac {1} {Z - E_r - \omega_q - \omega_{q'}} } \right )
\ F_{Y^*}(I) \; ,
\end{eqnarray}
with
$F_{Y^*} = \frac {1} {2} (3 - \vec \tau_1 \cdot \vec \tau_2)$.
By $p (p')$ we denote the four--momentum of the ingoing (outgoing)
nucleon, by $q (q')$ the four-momentum of the ingoing (outgoing) kaon,
whereas $p_r$ stands for the four--momentum of the exchanged hyperon.
We choose $p^0_r = \epsilon_N - \epsilon_K$ with
$\epsilon_N \equiv \frac {s + m^2_N - m^2_K} {2 \sqrt s}$,
$\epsilon_K \equiv \frac {s - m^2_N + m^2_K} {2 \sqrt s}$.

\begin{figure}
\caption{Contributions to $KN$ scattering. (a),(b): diagrams included
in Ref.\protect{\cite{Bue}}; (c): correlated $2 \pi$ exchange, which
was parametrized by (a) in Ref.\protect{\cite{Bue}}.}
\end{figure}

\begin{figure}
\caption{Model for the $N \bar N \rightarrow \pi \pi, K \bar K$
transition potentials.}
\end{figure}

\begin{figure}
\caption{Driving terms building up the coupled channels
($\pi \pi, K \bar K$) amplitude.}
\end{figure}

\begin{figure}
\caption{The $N \bar N \rightarrow K \bar K$ helicity amplitudes
$\tilde f^0_+$ (a) and $\tilde f^1_{\pm}$ (b), (c) as a function of $t$
in the pseudophysical region. The solid lines show the model result. The
dash--dotted line in (a) shows the result neglecting the
$N \bar N \rightarrow K \bar K$ transition potentials. The
vertical solid (dashed) line in (a) indicates the $\delta$ function at
$m_{\sigma} = 600(750)$ MeV representing sharp mass $\sigma$--exchange,
the vertical lines in (b) and (c) indicate the $\delta$ function at
$m_{\rho} = 769$ MeV representing sharp mass $\rho$--exchange.}
\end{figure}

\begin{figure}
\caption{Effective coupling constant $g_{\sigma}$ as a function of
$-t$. The dash--dotted line shows the $g_{\sigma}$ used in model I (with
$m_{\sigma} = 600$ MeV), the solid (dashed) line shows the result for
correlated $2 \pi$--exchange using $m_{\sigma} = 600 (750)$ MeV in the
parametrization.}
\end{figure}

\begin{figure}
\caption{Effective coupling constants for $\rho$--exchange as a function
of $-t$ (with $m_{\rho} = 769$ MeV). The dotted (double--dotted) line
shows $g_V$ ($g_T$) used in model I, the solid (dash--dotted) line shows
$g_V$ ($g_T$) for correlated $2 \pi$--exchange calculated with
Eq.(\protect{\ref{AB-rhof}}), the short dashed (long dashed) line $g_V$
($g_T$) for correlated $2 \pi$--exchange calculated with
Eq.(\protect{\ref{AB-rhoG}}).}
\end{figure}

\begin{figure}
\caption{$KN$ scattering phase shifts for $J = \frac {1}{2}$ and
$J = \frac {3}{2}$ as a function of the kaon laboratory momentum. The
solid (dash--dotted, dotted) line shows the result of model
I\negthinspace I A (model I\negthinspace I B, model I). Empirical
data is taken from \protect{\cite{Watts}} (empty circles),
\protect{\cite{Hashi}} (full circles) and \protect{\cite{Arndt}}
(empty squares).}
\end{figure}

\begin{figure}
\caption{The same as in Fig.~7 for $KN$ (I=0,1) elastic cross sections.
Experimental data is taken from \protect{\cite{Bowen}} (full circles),
\protect{\cite{Bugg}} (empty circles), \protect{\cite{Hashi}}
(empty squares).}
\end{figure}

\begin{figure}
\caption{The same as in Fig.~7 for $K^+p$ differential cross sections.
Experimental data is taken from \protect {\cite{Charles}}.}
\end{figure}

\begin{figure}
\caption{The same as in Fig.~7 for $K^+n$ differential cross sections.
Experimental data is taken from \protect{\cite{Giaco}}.}
\end{figure}

\begin{figure}
\caption{The same as in Fig.~7 for $K^+p$ and $K^+n$ polarizations.
Experimental data is taken from \protect{\cite{Lovett}} ($K^+p$) and
\protect{\cite{Robert}} ($K^+n$).}
\end{figure}

\begin{table}
\caption{Vertex parameters used in model I}
\begin{tabular}{cccccc}
Process & Exch. part. & $M_r$ or $m_r \, ^{a)}$
 & $g_1g_2 /4 \pi \, ^{b)}$
 & $\Lambda_1 \, ^{c)}$  & $\Lambda_2 \, ^{c)}$   \\
        &         & [$MeV$]   & [$f_1/g_1$]
 & [$GeV$] & [$GeV$]  \\
\hline
$K N \rightarrow KN$ & $\sigma$       & \phantom{1}600\phantom{.03}
 &\phantom{--4}1.300\phantom{[6.1]}
 & 1.7 & 1.5  \\
                     & $\sigma_{rep}$ & 1200\phantom{.03}
 & --40\phantom{.000}\phantom{[6.1]}
 & 1.5 & 1.5  \\
                     & $\omega$       & \phantom{1}782.6\phantom{3}
 &\phantom{--4}2.318 [0]\phantom{.1}
 & 1.5 & 1.5  \\
                     & $\rho$         & \phantom{1}769\phantom{.03}
 &\phantom{--4}0.773[6.1]
 & 1.4 & 1.6  \\
                     & $\Lambda$      & 1116\phantom{.03}
 &\phantom{--4}0.905\phantom{[6.1]}
 & 4.1 & 4.1  \\
                     & $\Sigma$       & 1193\phantom{.03}
 &\phantom{--4}0.031\phantom{[6.1]}
 & 4.1 & 4.1  \\
                     & $Y^*$          & 1385\phantom{.03}
 & \phantom{--4}0.037\phantom{[6.1]}
 & 1.8 & 1.8  \\
$K N \rightarrow K^* N$ & $\pi$          & \phantom{1}138.03
 &\phantom{--4}3.197\phantom{[6.1]}
 & 1.3 & 0.8  \\
                        & $\rho$         & \phantom{1}769\phantom{.03}
 &\phantom{--4}0.773[6.1]
 & 1.4 & 1.0  \\
$K N \rightarrow K^* \Delta$ & $\pi$          & \phantom{1}138.03
 &\phantom{--4}0.506\phantom{[6.1]}
 & 1.2 & 0.8  \\
                          & $\rho$         & \phantom{1}769\phantom{.03}
 &\phantom{--4}4.839\phantom{[6.1]}
 & 1.3 & 1.0  \\
$KN \rightarrow K \Delta$ & $\rho$         & \phantom{1}769\phantom{.03}
 &\phantom{--4}4.839\phantom{[6.1]}
 & 1.3 & 1.6
\end{tabular}
\vskip0.5cm
\noindent
\begin{flushleft}
$^{a)}$ Mass of exchanged particle. \\
$^{b)}$ Product of coupling constants [ratio of tensor to vector
        coupling]. \\
$^{c)}$ Cutoff mass. \\
\end{flushleft}
\end{table}

\begin{table}
\caption{Vertex parameters used in the models (I\negthinspace I A,B)
with correlated $2 \pi$--exchange$^{d,e)}$}
\begin{tabular}{cccccc}
Process & Exch. part. & $M_r$ or $m_r \, ^{a)}$
 & $g_1g_2 /4 \pi \, ^{b)}$
 & $\Lambda_1 \, ^{d)}$  & $\Lambda_2 \, ^{c)}$   \\
        &         & [$MeV$]   & [$f_1/g_1$]
 & [$GeV$] & [$GeV$]  \\
\hline
$K N \rightarrow KN$ & $\sigma_{rep}$ & 1600 (1200)
 & \phantom{(}--40 (--45)\phantom{[6.1]}
 & 2.1 (1.5)           & 2.1 (1.5)          \\
                     & $\omega$       & \phantom{1}782.6\phantom{3}
 &\phantom{--4}2.318[0]\phantom{.1}
 & 1.5 \phantom{(1.5)} & 1.5  \phantom{(1.5)} \\
                     & $\Lambda$      & 1116\phantom{.03}
 &\phantom{--4}0.905\phantom{[6.1]}
 & 3.5 (4.1)           & 5.0 (4.1)  \\
                     & $\Sigma$       & 1193\phantom{.03}
 &\phantom{--4}0.031\phantom{[6.1]}
 & 5.0 (4.1)           & 5.0 (4.1)  \\
                     & $Y^*$          & 1385\phantom{.03}
 &\phantom{--4}0.037\phantom{[6.1]}
 & 2.4 (1.8)           & 2.4 (1.8)  \\
$K N \rightarrow K^* N$ & $\pi$          & \phantom{1}138.03
 &\phantom{--4}3.197\phantom{[6.1]}
 & 1.3 \phantom{(1.3)} & 1.3 (0.8)  \\
                        & $\rho$         & \phantom{1}769\phantom{.03}
 &\phantom{--4}0.773[6.1]
 & 1.4 \phantom{(1.4)} & 1.1 (1.0)  \\
$K N \rightarrow K^* \Delta$ & $\pi$          & \phantom{1}138.03
 &\phantom{--4}0.506\phantom{[6.1]}
 & 1.2 \phantom{(1.2)} & 1.3 (0.8)  \\
                          & $\rho$         & \phantom{1}769\phantom{.03}
 &\phantom{--4}4.839\phantom{[6.1]}
 & 1.8 (1.6) & 1.1 (1.0)  \\
$KN \rightarrow K \Delta$ & $\rho$         & \phantom{1}769\phantom{.03}
 &\phantom{--4}4.839\phantom{[6.1]}
 & 1.8 (1.6) & 1.5 (1.4)
\end{tabular}
\vskip0.5cm
\noindent
\begin{flushleft}
$^{a)}$ Mass of exchanged particle. \\
$^{b)}$ Product of coupling constants [ratio of tensor to vector
        coupling]. \\
$^{c)}$ Cutoff mass. \\
$^{d)}$ We used for correlated $2 \pi$--exchange a cutoff
        $\Lambda_{\sigma,\rho}$ (Eq.(\ref{formfac})) of 1.85 GeV for
        method A \hspace{1cm}
\phantom{$^{d)}$} resp. 2.4 GeV for method B. \\
$^{e)}$ In case the parameters differ for model I\negthinspace I
        A, B, the numbers for B are given in \hspace{2cm}
\phantom{$^{e)}$} parentheses. \\
\end{flushleft}
\end{table}

\begin{table}
\caption{Low energy parameters}
\begin{tabular}{cccc}
                  & $a^0_{\frac{1}{2} S} [fm]$ &
                   $a^1_{\frac{1}{2} S} [fm]$ &
                   $r^1_{\frac{1}{2} S} [fm]$  \\[0.2cm]
\hline
 experiment$^{a)}$
& 0.03 $\pm$ 0.15 &
 -- 0.30 $\pm$ 0.03  &
    0.43 $\pm$ 0.22  \\
 model I\phantom {\negthinspace I A} & \phantom{--} 0.057 &
                -- 0.316 &
                   0.373  \\
 model I\negthinspace I A            & \phantom{--} 0.038 &
                -- 0.304 &
                   0.261  \\
 model I\negthinspace I B            &  -- 0.080 &
                -- 0.333 &
                   0.130  \\
\end{tabular}
\vskip0.5cm
\noindent
\begin{flushleft}
$^{a)}$ Empirical data is taken from Refs.\protect{\cite{ArRo,DoWa}}.\\
\end{flushleft}
\end{table}

\end {document}